\documentclass[aps,prl,showpacs,twocolumn,superscriptaddress,preprintnumbers,amsmath,amssymb]{revtex4}
\usepackage{graphicx} 
\usepackage{dcolumn}  
\usepackage{subfigure}

\renewcommand{\arraystretch}{1.1}

\newcommand{\ra}{\rightarrow}
\newcommand{\lam}{\Lambda}
\newcommand{\lamc}{\Lambda_c^+}
\newcommand{\sigc}{\Sigma_c(2455)}
\newcommand{\ks}{K^0_S}
\newcommand{\kl}{K^0_L}
\newcommand{\lamcst}{\Lambda_c(2880)^+}
\newcommand{\mnERR}{515.4 {^{+3.2}_{-3.1}} {^{+2.1}_{-6.0}}\,\mev}
\newcommand{\mppERR}{514.5 {^{+3.4}_{-3.1}} {^{+2.8}_{-4.9}}\,\mev}
\newcommand{\mpERR}{505.4 {^{+5.8}_{-4.6}} {^{+12.4}_{-\phantom{1}2.0}}\,\mev}
\newcommand{\gnERR}{61 {^{+18}_{-13}} {^{+22}_{-13}}\,\mathrm{MeV}}
\newcommand{\gppERR}{75 {^{+18}_{-13}} {^{+12}_{-11}}\,\mathrm{MeV}}
\newcommand{\gpERR}{62 {^{+37}_{-23}} {^{+52}_{-38}}\,\mathrm{MeV}}
\newcommand{\mnERRo}{515.4 {^{+3.2}_{-3.1}} {^{+2.1}_{-6.0}}}
\newcommand{\mppERRo}{514.5 {^{+3.4}_{-3.1}} {^{+2.8}_{-4.9}}}
\newcommand{\mpERRo}{505.4 {^{+5.8}_{-4.6}} {^{+12.4}_{-\phantom{1}2.0}}}
\newcommand{\gnERRo}{61 {^{+18}_{-13}} {^{+22}_{-13}}}
\newcommand{\gppERRo}{75 {^{+18}_{-13}} {^{+12}_{-11}}}
\newcommand{\gpERRo}{62 {^{+37}_{-23}} {^{+52}_{-38}}}
\newcommand{\mev}{\mathrm{MeV}/c^2}
\newcommand{\gev}{\mathrm{GeV}/c^2}
\newcommand{\dm}{\Delta M(\Lambda_c^+ \pi)}

\newcommand{\dmmp}{\Delta M(\Lambda_c^+ \pi^\mp)}
\newcommand{\dmm}{\Delta M(\Lambda_c^+ \pi^-)}
\newcommand{\dmp}{\Delta M(\Lambda_c^+ \pi^+)}
\newcommand{\thdec}{\theta_{\text{dec}}}
\newcommand{\costh}{\cos \thdec}
\newcommand{\like}{\mathcal{L}}
\newcommand{\br}{\mathcal{B}}
\newcommand{\scstn}{\Sigma_c(2800)^0}
\newcommand{\scstp}{\Sigma_c(2800)^+}
\newcommand{\scstpp}{\Sigma_c(2800)^{++}}

\begin{document}

\title{Observation of an Isotriplet of Excited Charmed Baryons 
Decaying to $\lamc \pi$ }
\date{\today}

\begin{abstract}
\noindent
We report the observation of an isotriplet of excited charmed baryons,
decaying into $\lamc \pi^-$, $\lamc \pi^0$ and $\lamc \pi^+$. 
We measure the mass differences $M(\lamc \pi)-M(\lamc)$ and widths
to be $\mnERR$, $\gnERR$ for the neutral state; $\mpERR$, 
$\gpERR$ for the charged state and $\mppERR$, $\gppERR$ for the doubly
charged state, 
where the uncertainties are statistical and systematic, respectively. 
These
results are obtained from a $281\,{\rm fb}^{-1}$ data sample collected
with the Belle detector near the $\Upsilon(4S)$ resonance, at the KEKB
asymmetric energy $e^+ e^-$ collider.

\end{abstract}


\affiliation{Budker Institute of Nuclear Physics, Novosibirsk}
\affiliation{Chiba University, Chiba}
\affiliation{Chonnam National University, Kwangju}
\affiliation{University of Cincinnati, Cincinnati, Ohio 45221}
\affiliation{Gyeongsang National University, Chinju}
\affiliation{University of Hawaii, Honolulu, Hawaii 96822}
\affiliation{High Energy Accelerator Research Organization (KEK), Tsukuba}
\affiliation{Hiroshima Institute of Technology, Hiroshima}
\affiliation{Institute of High Energy Physics, Chinese Academy of Sciences, Beijing}
\affiliation{Institute of High Energy Physics, Vienna}
\affiliation{Institute for Theoretical and Experimental Physics, Moscow}
\affiliation{J. Stefan Institute, Ljubljana}
\affiliation{Kanagawa University, Yokohama}
\affiliation{Korea University, Seoul}
\affiliation{Kyungpook National University, Taegu}
\affiliation{Swiss Federal Institute of Technology of Lausanne, EPFL, Lausanne}
\affiliation{University of Ljubljana, Ljubljana}
\affiliation{University of Maribor, Maribor}
\affiliation{University of Melbourne, Victoria}
\affiliation{Nagoya University, Nagoya}
\affiliation{Nara Women's University, Nara}
\affiliation{National Central University, Chung-li}
\affiliation{National United University, Miao Li}
\affiliation{Department of Physics, National Taiwan University, Taipei}
\affiliation{H. Niewodniczanski Institute of Nuclear Physics, Krakow}
\affiliation{Nihon Dental College, Niigata}
\affiliation{Niigata University, Niigata}
\affiliation{Osaka City University, Osaka}
\affiliation{Osaka University, Osaka}
\affiliation{Panjab University, Chandigarh}
\affiliation{Peking University, Beijing}
\affiliation{Princeton University, Princeton, New Jersey 08545}
\affiliation{University of Science and Technology of China, Hefei}
\affiliation{Seoul National University, Seoul}
\affiliation{Sungkyunkwan University, Suwon}
\affiliation{University of Sydney, Sydney NSW}
\affiliation{Tata Institute of Fundamental Research, Bombay}
\affiliation{Toho University, Funabashi}
\affiliation{Tohoku Gakuin University, Tagajo}
\affiliation{Tohoku University, Sendai}
\affiliation{Department of Physics, University of Tokyo, Tokyo}
\affiliation{Tokyo Institute of Technology, Tokyo}
\affiliation{Tokyo Metropolitan University, Tokyo}
\affiliation{University of Tsukuba, Tsukuba}
\affiliation{Virginia Polytechnic Institute and State University, Blacksburg, Virginia 24061}
\affiliation{Yonsei University, Seoul}
  \author{R.~Mizuk}\affiliation{Institute for Theoretical and Experimental Physics, Moscow} 
  \author{K.~Abe}\affiliation{High Energy Accelerator Research Organization (KEK), Tsukuba} 
  \author{K.~Abe}\affiliation{Tohoku Gakuin University, Tagajo} 
  \author{H.~Aihara}\affiliation{Department of Physics, University of Tokyo, Tokyo} 
  \author{M.~Akatsu}\affiliation{Nagoya University, Nagoya} 
  \author{Y.~Asano}\affiliation{University of Tsukuba, Tsukuba} 
  \author{V.~Aulchenko}\affiliation{Budker Institute of Nuclear Physics, Novosibirsk} 
  \author{T.~Aushev}\affiliation{Institute for Theoretical and Experimental Physics, Moscow} 
  \author{A.~M.~Bakich}\affiliation{University of Sydney, Sydney NSW} 
  \author{V.~Balagura}\affiliation{Institute for Theoretical and Experimental Physics, Moscow} 
% ...
\collaboration{The Belle Collaboration}
\maketitle

{\renewcommand{\thefootnote}{\fnsymbol{footnote}}}
\setcounter{footnote}{0}


Charmed baryon spectroscopy provides an excellent laboratory to study the
dynamics of a light diquark in the environment of a heavy quark,
allowing the predictions of different theoretical approaches to be
tested~\cite{motiv}. The baryons containing one $c$ quark and two
light ($u$ or $d$) quarks are denoted $\Lambda_c$ and $\Sigma_c$ for
states with isospin zero and one, respectively. 
All known excited charmed baryons decay into $\lamc\pi$ and
$\lamc\pi\pi$ final states. 
There are four excited charmed baryons observed
in the $\lamc\pi^+\pi^-$ final state~\cite{pdg}; the lower two are
identified as orbital excitations of $\lamc$ while the
upper two are not yet identified. In the
$\lamc\pi$ final state only the $\Sigma_c(2455)$ ground state and
the $\Sigma_c(2520)$ spin excitation have been observed so far, while the
orbital excitations of the $\Sigma_c$ remain to be found. 
In this Letter we present the results of a search for
new states decaying into a $\lamc$ baryon and                                   
a charged or neutral pion. 
This study is performed using a data sample of $253
\,\mathrm{fb}^{-1}$ collected at the $\Upsilon(4S)$ resonance and
$28\,\mathrm{fb}^{-1}$ at an energy $60\,{\mathrm{MeV}}$ below the
resonance. The data were collected with the Belle
detector~\cite{BELLE_DETECTOR} at the KEKB asymmetric energy $e^+ e^-$
storage rings~\cite{KEKB}.

The Belle detector is a large-solid-angle magnetic spectrometer that
consists of a silicon vertex detector (SVD), a 50-layer
cylindrical drift chamber (CDC), an array of aerogel threshold
Cherenkov counters (ACC), a barrel-like array of time-of-flight
scintillation counters (TOF), and an array of CsI(Tl) crystals (ECL)
located inside a superconducting solenoidal coil that produces a 1.5T
magnetic field. An iron flux return located outside the coil is
instrumented to detect muons and $\kl$ mesons (KLM).
Two different inner detector configurations were used. 
For the first sample of $155\,\mathrm{fb}^{-1}$, a $2.0\,$cm radius
beampipe and a 3-layer silicon vertex detector were used;
for the latter sample of $126\,\mathrm{fb}^{-1}$, a $1.5\,$cm radius
beampipe and a 4-layer silicon vertex detector and a small-cell inner
drift chamber were used~\cite{new_svd}. 
We use a GEANT based Monte-Carlo (MC) simulation to model the response of
the detector and to determine its acceptance. Signal MC events are
produced with run dependent conditions and correspond to relative
luminosities of different running periods.


$\lamc$ baryons are reconstructed using the $p K^-\pi^+$ decay mode 
(the inclusion of charge conjugate modes 
is implied throughout this Letter). 
Charged
hadron candidates are required to originate from the
vicinity of the run-averaged interaction point. 
For charged particle identification (PID), 
the combined information from CDC ($dE/dx$), TOF and ACC 
is used. 
Protons, charged kaons and pions 
are selected with PID criteria 
that have efficiencies of 83\%, 84\% and 90\%, respectively 
(the PID criteria are not applied for 
pions originating from $\lamc$ decays).
The PID criteria reduce the background to 3\%, 13\% and 53\%,
respectively. 
In addition we remove charged hadron candidates if they are
consistent with being electrons based on the ECL, CDC and ACC
information. 
We define the signal window around the $\lamc$ mass 
to be $\pm 8\,\mev$ which corresponds to an efficiency of about $90\%$
($1.6\sigma$). 

A pair of calorimeter showers with an invariant mass within $10\,\mev$
($1.6\sigma$) of the nominal $\pi^0$ mass is considered as a $\pi^0$
candidate. An energy of at least 50$\,\mathrm{MeV}$ is required for
each shower. 

To reduce the combinatorial background in $\lamc$
and $\lamc\pi$ resonance reconstruction, we impose a
requirement on the scaled momentum
$x_p \equiv p^\ast/p^\ast_{\text{max}}$, where $p^\ast$ is the momentum 
of the charmed baryon candidate in the center of mass (c.m.) frame, 
and $p^\ast_{\text{max}} \equiv \sqrt{E^{\ast 2}_{\rm beam}-M^2}$; 
$E^\ast_{\rm beam}$ is the beam energy in the c.m. frame, 
and $M$ is the mass of the candidate. 
To allow a comparison of our $\lamc$ sample with that of other
experiments and to demonstrate its high purity,
we apply a $x_p>0.5$ requirement on $\lamc$ candidates. 
The $\lamc$ yield with this requirement is  
$(516\pm 2)\times 10^3$ and the signal-to-background ratio is 2.3. 

We combine $\lamc$ candidates with the remaining pion candidates in the event. 
The $x_p$ requirement on the $\lamc$ candidate is released, and 
a $x_p>0.7$ requirement on the $\lamc\pi$ pair is applied. 
The tight $x_p$ cut is justified by the hardness of the momentum
spectra of known excited charmed baryons. 
To further suppress the combinatorial
background from low momentum pions we require the decay angle $\thdec$
to satisfy $\costh > -0.4$.  $\thdec$ is defined as
the angle between the $\pi$ momentum measured in the rest frame of
the $\lamc \pi$ system, and the boost direction of the $\lamc
\pi$ system in the c.m. frame. The requirement $\costh > -0.4$ is
chosen assuming a flat $\costh$ distribution
for the signal.

\begin{figure*}[bth]
\centering
\begin{picture}(550,200)
\put(3,130){\rotatebox{90}{$N / 10~\mev$}} 
\put(65,170){$\lamc\pi^-$} 
\put(215,170){$\lamc\pi^0$} 
\put(365,170){$\lamc\pi^+$} 
\put(380,8){$M(\lamc \pi)-M(\lamc),~\gev$} 
\put(-5,-10){\includegraphics[width=1.07\textwidth]{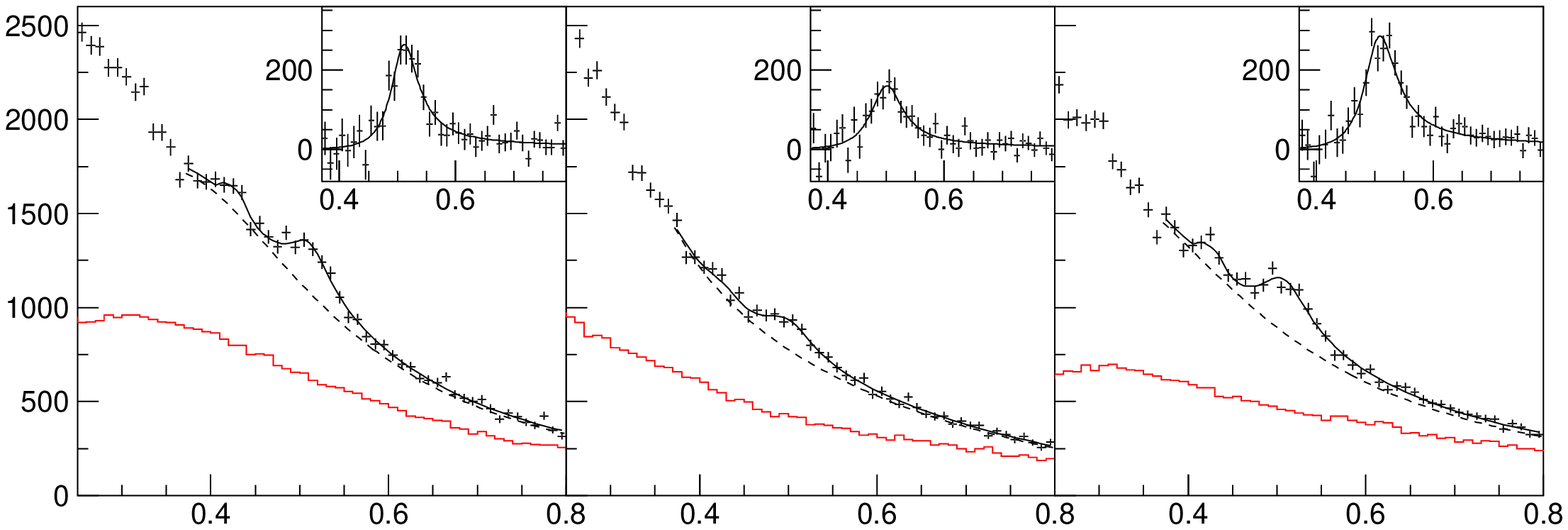}}
\end{picture}
\caption{$M(\lamc \pi) - M(\lamc)$ distributions of the selected
  $\lamc \pi^-$ (left), $\lamc \pi^0$ (middle) and $\lamc \pi^+$
  (right) combinations. Data from the $\lamc$ signal 
  window (points with error bars) and normalized sidebands 
  (histograms) are shown, together with the fits described in the text
  (solid curves) and their combinatorial background components (dashed).
  The insets show the background subtracted distributions in the signal
  region (points with error bars) with the signal component from the fit
  superimposed.
}
\label{m_sigc2}
\end{figure*}

Fig.~\ref{m_sigc2} 
shows distributions of the mass difference
$\dm \equiv M(\lamc \pi) - M(\lamc)$ for the 
$\lamc\pi^-$, $\lamc\pi^0$ and $\lamc\pi^+$ combinations 
in the region above the $\Sigma_c(2455)$ and $\Sigma_c(2520)$
resonances.
All the distributions 
show enhancements near $0.51\,\gev$, which we interpret as
signals of new excited charmed baryons, 
forming an isotriplet. 
The new baryons are hereafter denoted the
$\Sigma_c(2800)^0$, $\Sigma_c(2800)^+$ and $\Sigma_c(2800)^{++}$ for
the three final states, respectively. 
Scaled $\lamc$ sidebands, which are also shown in Fig.~\ref{m_sigc2}, 
exhibit featureless $\Delta M$ distributions.
We also check the
$\dm$ spectra for $e^+ e^- \to c\bar{c}$ MC events, and find no
enhancement in our signal region.

\begin{figure}[tbh]
\centering
\begin{picture}(550,210)
\put(10,130){\rotatebox{90}{$N / 20~\mev$}} 
\put(160,180){$\lamc\pi^+$} 
\put(160,100){$\lamc\pi^-$} 
\put(100,8){$M(\lamc\pi^\pm_{\rm rem})-M(\lamc),~\gev$} 
\put(10,-10){\includegraphics[width=0.50\textwidth]{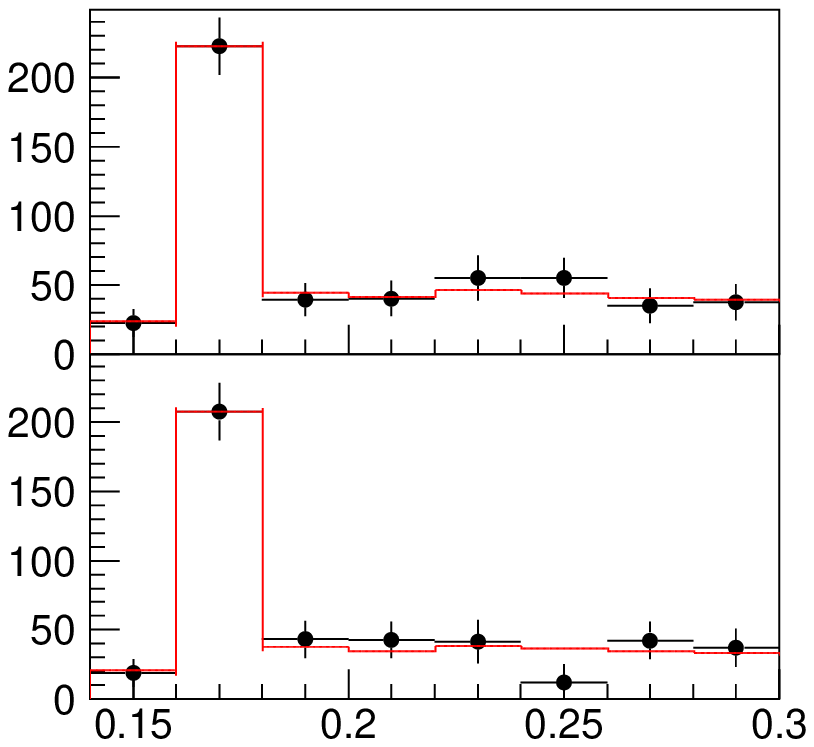}}
\end{picture}
\caption{Yield of $\lamcst$ {\emph vs.} 
$\Delta M(\lamc\pi^+_{\rm rem})$ (top) and 
$\Delta M(\lamc\pi^-_{\rm rem})$ (bottom).
The peaks are due to intermediate $\sigc^{++}$ and $\sigc^{0}$ states,
respectively, in $\lamcst$ decay: see the text.}
\label{feedd}
\end{figure}

The enhancement near $\Delta M = 0.43\,\gev$ 
in the $\dmm$ and $\dmp$ spectra 
is attributed to feed-down from the decay 
$\lamcst \to \lamc \pi^+ \pi^-$. 
The $\lamcst$ resonance was observed by
CLEO~\cite{cleo_lamc2880} in the $\lamc\pi^+\pi^-$ final state;
30\% of decays proceed via an intermediate $\sigc^0$
or $\sigc^{++}$. From a MC study we find that if $\lamc\pi^\pm$
pairs are produced from intermediate $\sigc^{++/0}$, then the $\dmmp$ 
spectrum is peaked around $0.43\,\gev$.  To
determine the yield of the feed-down we reconstruct the 
$\lamcst\ra\lamc\pi^+\pi^-$ decays: selected $\lamc\pi^\mp$
pairs are combined with all remaining pions $\pi^\pm_{\rm rem}$ in
the event.
We observe clear peaks of $\Lambda_c(2880)^+$ and $\Lambda_c(2765)^+$, 
consistent with the observation of these states by CLEO.
We then fit the
$\Delta M(\lamc\pi^\mp\pi^\pm_{\rm rem}) \equiv 
        M(\lamc\pi^\mp\pi^\pm_{\rm rem}) - M(\lamc)$
spectra to obtain the $\lamcst$ yield in bins of
$\Delta M(\lamc\pi^+_{\rm rem})$ and $\Delta M(\lamc\pi^-_{\rm rem})$.
The results of the fits are shown in Fig.~\ref{feedd}.
Each distribution shows a peak in the second bin 
due to an intermediate $\sigc$ state.  The fitting
function, shown in Fig.~\ref{feedd}, 
includes both resonant and non-resonant contributions and is
determined from the MC simulation. The result of this fit is used
to determine the $\sigc^{++/0}$ fractions in $\lamcst$ decay, and thus
the shape of the $\lamcst$ feed-down to the $\lamc\pi^\mp$ distributions.
In this calculation, we correct the feed-down normalization 
for the efficiency of $\pi^\pm_{\rm rem}$ reconstruction. 

For the $\lamc\pi^0$ final state we expect a feed-down from the
$\lamcst\to\lamc\pi^0\pi^0$ decay. 
If the $\lamcst$ isospin is zero, 
then the following relations are valid:
$\br(\lamcst\to\lamc\pi^0\pi^0) = 
0.5\cdot\br(\lamcst\to\lamc\pi^+\pi^-)$ and 
$\br(\lamcst\to\sigc^+\pi^0) = 
\br(\lamcst\to\sigc^{++}\pi^-) = 
\br(\lamcst\to\sigc^0\pi^+)$.
We do not observe the $\lamcst\to\lamc\pi^0\pi^0$ decay 
due to the lower reconstruction efficiency 
for $\pi^0$, compared to $\pi^\pm$ 
(the expected signal yield is about 100 events,
while the square root of the background is 110). 
Therefore the shape and normalization of the
$\lamcst\to\lamc\pi^0\pi^0$ feed-down 
is determined based on the 
$\lamcst\to\lamc\pi^+\pi^-$ feed-down 
and the above isospin relations.
We take into account the differences 
in the reconstruction efficiencies and
mass resolutions for the $\lamc\pi^0$ 
and $\lamc\pi^\mp$ final states. 
If the $\lamcst$ isospin is one, then the $\lamcst\to\lamc\pi^0\pi^0$
decay is forbidden. This possibility is taken into account as a
systematic uncertainty. 

We perform a fit to the $\lamc\pi$ mass spectra of
Fig.~\ref{m_sigc2} to extract the parameters and yields of
the $\Sigma_c(2800)^0$, $\Sigma_c(2800)^+$ and $\Sigma_c(2800)^{++}$. 
The fitting function is a sum of three components: 
signal, feed-down and combinatorial background functions.
We tentatively identify the $\Sigma_c(2800)$ states as $\Sigma_{c2}$ baryons,
decaying to $\lamc\pi$ in D-wave, so the 
signal is parameterized by a relativistic D-wave Breit-Wigner function 
(Blatt-Weisskopf form factors~\cite{blatt_weisskopf} are included),
convolved with the detector resolution of $2\,\mev$ ($7\,\mev$)
for final states containing only charged (one neutral) pions.
The shape and the normalization of the feed-down from $\lamcst$ 
is fixed as described above. 
The background is parameterized by an inverse third
order polynomial ($1/\{C_0+C_1x+C_2x^2+C_3x^3\}$, where the $C_i$ are
floating parameters). 
The fit interval starts at $0.37\,\gev$ since at lower mass a feed-down
from $\Lambda_c(2765)^+\to\lamc\pi^+\pi^-$ \cite{cleo_lamc2880} is expected. 
The fits are shown in Fig.~\ref{m_sigc2},
and their results are summarized in Table~\ref{table_fit}.
The signal yield is defined as the integral of the Breit-Wigner function 
over the mass interval $0.34\,\gev <\Delta M<0.69\,\gev$ ($\sim 2.5 \Gamma$).
\begin{table}[htb]
\caption{Signal yield, mass and width for $\scstn$, $\scstp$
  and $\scstpp$. The first uncertainty is statistical, the second one
  is systematic.} 
\label{table_fit}
\renewcommand{\arraystretch}{1.2}
\begin{ruledtabular}
\begin{tabular}{lccc}
  State		& Yield $/10^3$ & $\Delta M,~{\rm MeV}/c^2$  & $\Gamma,~{\rm MeV}$\\
  \hline
  $\scstn$	& ${2.24^{+0.79}_{-0.55}}{^{+1.03}_{-0.50}}$ & $\mnERRo$ & $\gnERRo$ \\
  $\scstp$	& ${1.54^{+1.05}_{-0.57}}{^{+1.40}_{-0.88}}$ & $\mpERRo$ & $\gpERRo$ \\
  $\scstpp$	& ${2.81^{+0.82}_{-0.60}}{^{+0.71}_{-0.49}}$ & $\mppERRo$& $\gppERRo$\\
\end{tabular}
\end{ruledtabular}
\end{table}

The signal significances are 8.6, 6.2 and 10.0 standard
deviations for $\Sigma_c(2800)^0$, $\Sigma_c(2800)^+$ 
and $\Sigma_c(2800)^{++}$, respectively. 
The significance is defined as
$\sqrt{-2\ln{(\like_0/\like_{\text{max}})}}$,
where $\like_0$ and $\like_{\text{max}}$ are the
likelihood values returned by fits with the signal yield fixed at zero
and the best fit values respectively.

To estimate the systematic uncertainty on the results of the fit 
we vary the signal parameterization,
using S-wave and P-wave Breit-Wigner functions. 
We vary the interval in $\Delta M$ used for the fit and 
the background parameterization, using polynomials and inverse
polynomials of different orders,
a function $(C_1+C_2x)
\cdot \exp{(C_3x+C_4x^2)}$, where the $C_i$ are floating
parameters, and these functions plus the normalized $\lamc$ sidebands. 
We vary 
the normalization of the $\lamcst$ feed-down by $\pm 2\sigma$ and  
in the $\lamc\pi^0$ case we also consider the possibility of zero
feed-down. 
We tighten the $x_p$ cut to 0.75 
and we vary the $\costh$ cut from $-0.5$ to $-0.3$.
In each case we take the largest positive and negative variation in the 
fitted parameters as the systematic uncertainty from this source;
each term is then added in quadrature to give the total systematic uncertainty.
The dominant contribution originates from the variation of the
interval used in the fit and the background parameterization. 
The parameters for the observed states with statistical and systematic
uncertainties are summarized in Table~\ref{table_fit}. 
For all background parameterizations the signal significance
exceeds 5.3$\sigma$ in all final states. 
As a cross-check, we repeat the analysis using the 
$\lamc\to p \ks$ and $\lam \pi^+$ decay modes, 
and find consistent results. 
(We do not perform an average since the relative yields are low 
and the branching ratios of these modes relative to 
$\lamc\to p K^- \pi^+$ are poorly known).
We also check that the observed
$\Sigma_c(2800)$ signals are not feed-downs from unknown
resonances decaying to $\lamc\pi\pi$, by combining
$\lamc\pi^\mp$ pairs from the $\Sigma_c(2800)$ region with
all remaining $\pi^\pm$ in the event. The resulting 
$\Delta M (\lamc\pi^+\pi^-)$ spectra exhibit no structure.

To determine the efficiency of the $\costh>-0.4$ requirement, 
we assume that the $\costh$ distribution is symmetric about zero, 
as required by the conservation of P-parity in strong decays. 
We check that the observed distributions are compatible with this 
assumption. 
We fit the $\Delta M(\lamc\pi)$ spectra
in the $0.4 < \costh \leq 1.0$ interval, fixing 
the signal parameters
to the values obtained above and assuming the same background
parameterization. The $\lamcst$ feed-down normalization is 
determined for the selected  $\costh$ interval. The reconstruction
efficiency corrections are taken into account. 
The statistical uncertainty in the obtained signal yield is included
in the systematic uncertainty of the efficiency.

\begin{figure}[tbh]
\centering
\begin{picture}(550,280)
\put(10,240){\rotatebox{90}{$N / 0.1$}}
\put(90,250){$\lamc\pi^-$} 
\put(90,170){$\lamc\pi^0$} 
\put(90,95){$\lamc\pi^+$} 
\put(215,8){\large $x_p$} 
\put(10,-10){\includegraphics[width=0.50\textwidth]{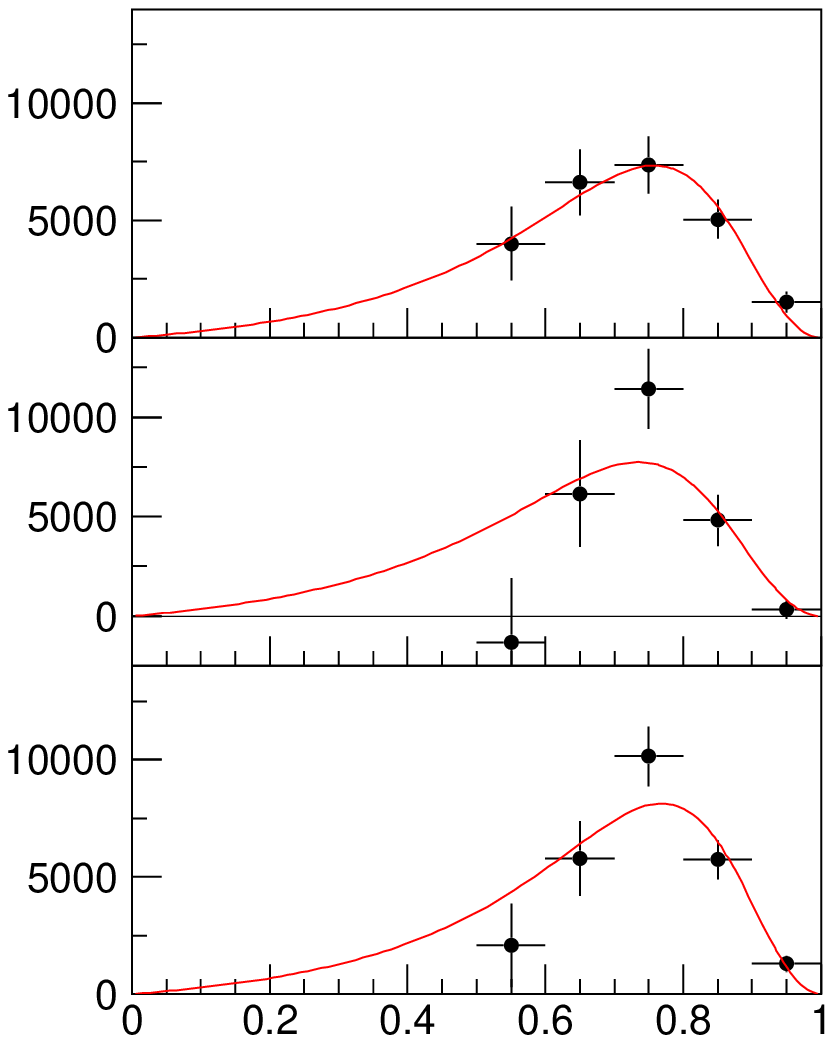}}
\end{picture}
\caption{Spectra of scaled momentum $x_p$ for $\scstn$
(top), for $\scstp$ (middle) and for $\scstpp$ (bottom) states.}
\label{xp_res}
\end{figure}
To find the efficiency of the $x_p>0.7$ requirement, 
we consider an extended interval $x_p>0.5$ and fit the
$\Delta M(\lamc\pi)$ spectra in $x_p$ bins. 
The $\lamcst$ feed-down normalization is 
determined in each $x_p$ bin separately.
The obtained efficiency-corrected $x_p$ spectra are shown in
Fig.~\ref{xp_res}; the curves show fits using the
Peterson~\cite{peterson} parameterization $dN/dx_p \sim x_p^{-1}
[1-1/x_p-\epsilon/(1-x_p)^2 ]^{-2}$.
We obtain $\epsilon = 0.078^{+0.021}_{-0.017}$ for 
$\scstn$, $\epsilon = 0.095^{+0.032}_{-0.024}$ for $\scstp$ and 
$\epsilon = 0.069^{+0.015}_{-0.012}$ for $\scstpp$; 
these values are similar
to other measurements for excited charmed baryons with nonzero
orbital angular momentum ~\cite{other_excited}.
The efficiency of the $x_p>0.7$ requirement is calculated using 
the $\epsilon$ values from the fits. 
The statistical uncertainty in $\epsilon$ is included
in the systematic uncertainty of the efficiency.

We calculate the product
$\sigma[e^+e^-\ra\Sigma_c(2800)X]\times
\br[\Sigma_c(2800)\ra\lamc\pi]$ to be
$(2.04{^{+0.72}_{-0.50}}{^{+0.97}_{-0.52}} \pm 0.53)\,\mathrm{pb}$
for $\scstn$, 
$(2.6{^{+1.8}_{-1.0}}{^{+2.4}_{-1.5}}\pm 0.7)\,\mathrm{pb}$ 
for $\scstp$ and 
$(2.36{^{+0.69}_{-0.50}}{^{+0.64}_{-0.47}}\pm 0.61)\,\mathrm{pb}$ 
for $\scstpp$; the first uncertainty is statistical, the second
is systematic, and the third is due to the uncertainty in
the $\lamc\to pK^-\pi^+$ branching fraction.

Theoretical models predict a rich spectrum of excited charmed
baryons in the vicinity of the observed states~\cite{copley}.  One of
the candidates is a $\Sigma_{c2}$ doublet with $J^P=3/2^-$ and
$5/2^-$, where the subscript 2 denotes the total angular momentum of
the light quark system.  The $\Sigma_{c2}$ baryon is expected to
decay principally into the $\lamc\pi$ final state in D-wave;
the predicted mass difference $\Delta M=500\,\mev$ is close to that observed here,
but the expected width $\Gamma \sim 15\,\mathrm{MeV}$~\cite{Pirjol} is smaller
than the one we observe.
However, we note that the $\Sigma_{c2}~(J^P=3/2^-)$ baryon can mix with the nearby
$\Sigma_{c1}~(J^P=3/2^-)$, which would produce a wider physical state.

In summary, we report the observation of an isotriplet of excited charmed baryons,
decaying into $\lamc \pi^-$, $\lamc \pi^0$ and $\lamc \pi^+$. 
We measure the mass differences $M(\lamc \pi)-M(\lamc)$ and widths
to be $\mnERR$, $\gnERR$ for the neutral state; $\mpERR$, 
$\gpERR$ for the charged state and $\mppERR$, $\gppERR$ for the doubly
charged state, 
where the uncertainties are statistical and systematic, respectively. 
We tentatively identify these states as members of the predicted 
$\Sigma_{c2},\, J^P = 3/2^-$ isospin triplet.

We thank the KEKB group for the excellent operation of the
accelerator, the KEK Cryogenics group for the efficient
operation of the solenoid, and the KEK computer group and
the NII for valuable computing and Super-SINET network
support.  We acknowledge support from MEXT and JSPS (Japan);
ARC and DEST (Australia); NSFC (contract No.~10175071,
China); DST (India); the BK21 program of MOEHRD and the CHEP
SRC program of KOSEF (Korea); KBN (contract No.~2P03B 01324,
Poland); MIST (Russia); MESS (Slovenia); Swiss NSF; NSC and MOE
(Taiwan); and DOE (USA).

\end{document}